\documentclass[runningheads]{llncs}
\usepackage[table,xcdraw]{xcolor}
\usepackage{graphicx}
\usepackage{amsmath}
\usepackage{amssymb}
\usepackage{multirow}
\usepackage{xcolor}
\usepackage{colortbl}
\usepackage{pgfplots}
\usepackage{layout}
\usepackage{amssymb}
\usepackage{bbding}
\usepackage{booktabs}
\usepackage{orcidlink}
\usepackage{diagbox}
\usepackage[misc]{ifsym}
\usepackage{makecell}

\usepackage[T1]{fontenc}

\usepackage{graphicx,verbatim}

\begin{document}
\title{
HWA-UNETR: Hierarchical Window Aggregate UNETR for 3D Multimodal Gastric Lesion Segmentation
}
\titlerunning{HWA-UNETR: Hierarchical Window Aggregate UNETR}

\author{
Jiaming Liang 
\inst{1}\orcidlink{0009-0000-6007-9759} \and
Lihuan Dai 
\inst{2}\orcidlink{0000-0002-0713-105X}
\and
Xiaoqi Sheng 
\inst{1}\orcidlink{0000-0002-2929-5805}
\and
Xiangguang Chen 
\inst{3}
\and
Chun Yao 
\inst{3}
\and
Guihua Tao 
\inst{4}
\and
Qibin Leng 
\inst{5}
\and
Hongmin Cai 
\inst{1}\orcidlink{0000-0002-2747-7234
}
\and  
Xi Zhong 
\inst{2}\orcidlink{0009-0007-0170-5482}\textsuperscript{(\Letter)}
}  

\authorrunning{Liang et al.}

\institute{
School of Computer Science and Engineering, School of Future Technology, South China University of Technology, Guangzhou, China 
\and 
Department of Medical Imaging, Guangzhou Institute of Cancer Research, the Affiliated Cancer Hospital, Guangzhou Medical University, Guangzhou, China
\email{zhongxi@gzhmu.edu.cn}
\and
Department of Radiology, Meizhou People’s Hospital, Meizhou, China
\and
School of Computer Science and Technology, Hainan University, Hainan, China 
\and 
Department of Oncology Institute, Guangzhou Institute of Cancer Research, the Affiliated Cancer Hospital, Guangzhou Medical University, Guangzhou, China
}

\maketitle              
\begin{abstract}
Multimodal medical image segmentation faces significant challenges in the context of gastric cancer lesion analysis. This clinical context is defined by the scarcity of independent multimodal datasets and the imperative to amalgamate inherently misaligned modalities. As a result, algorithms are constrained to train on approximate data and depend on application migration, leading to substantial resource expenditure and a potential decline in analysis accuracy. To address those challenges, we have made two major contributions: First, we publicly disseminate the GCM 2025 dataset, which serves as the first large-scale, open-source collection of gastric cancer multimodal MRI scans, featuring professionally annotated FS-T2W, CE-T1W, and ADC images from 500 patients. Second, we introduce HWA-UNETR, a novel 3D segmentation framework that employs an original HWA block with learnable window aggregation layers to establish dynamic feature correspondences between different modalities' anatomical structures, and leverages the innovative tri-orientated fusion mamba mechanism for context modeling and capturing long-range spatial dependencies. Extensive experiments on our GCM 2025 dataset and the publicly BraTS 2021 dataset validate the performance of our framework, demonstrating that the new approach surpasses existing methods by up to 1.68\% in the Dice score while maintaining solid robustness. The dataset and code are public via 
this 
\href{https://github.com/JeMing-creater/HWA-UNETR}{URL}.


\end{abstract}

\section{Introduction}
Gastric Cancer (GC) ranks as the fifth most common malignant neoplasm globally and the fourth leading cause of cancer-related mortality~\cite{sung2021global}, whose detection of lesions is a critical step in clinical treatment. Through sustained technical refinements, Magnetic Resonance Imaging (MRI) has evolved into a superior modality for gastrointestinal malignancy evaluation~\cite{borggreve2019imaging}, with its diagnostic ascendancy over Computed Tomography (CT) stemming from three cardinal virtues: multi-parametric soft tissue discrimination, absence of ionizing radiation, and synergistic incorporation of functional molecular imaging paradigms~\cite{singh2024treatment}. Recently, with the advancement of Deep Learning (DL) technologies, quantitative multimodal MRI scans have leveraged their inherent advantages to serve as an essential tool for detecting cancerous lesions~\cite{xing2024segmamba,liang2023agilenet,hatamizadeh2022swin}. 
More importantly, these technologies have significantly enhanced medical image segmentation methods~\cite{pang2023slim,liang2024comprehensive}, allowing for the systematic delineation of malignant tumor spatial distributions and precise quantification of lesion heterogeneity through the integration of multimodal data.

Existing DL methods are sensitive to imaging modalities and lesion characteristics, leading to segmentation performance that relies heavily on training data quality~\cite{pang2025online,bibars2023cross,ali2024review}. However, the scarcity of large-scale, high-quality multimodal
MRI datasets for GC necessitate reliance on generic medical imaging datasets and transfer learning~\cite{ma2024segment}, both of which are resource-intensive and reduce specificity. Furthermore, gastric cancer lesion detection involves multiple spatially misaligned MRI modalities, which pose two critical challenges for DL methods: (1) \textbf{aligning inherently misaligned imaging modalities and integrating complementary features}, and (2) \textbf{fully leveraging the complementary characteristics of each modality for multi-scale context modeling}.
Previous studies enhance segmentation via feature fusion~\cite{wenxuan2021transbts,zhou2019review}, yet clinical applicability remains limited due to reliance on overly optimistic data assumptions.
Additionally, improving model representational capacity has proven effective for segmentation tasks.
CNN-based~\cite{isensee2020nnu,ronneberger2015u} and Transformer-based methods~\cite{hatamizadeh2022swin,liang2024comprehensive,pang2023slim} improve contextual feature modeling, while Mamba-based methods~\cite{xing2024segmamba,gong2024nnmamba} optimize the modeling process by incorporating state space models (SSMs). Nonetheless, these predominantly focus on pixel-level predictions within single modalities, neglecting multimodal misalignment and complementarity. 
Recently,  H-DenseFormer~\cite{shi2023h} proposes a multi-path parallel embedding framework to enhance segmentation efficiency and accuracy in multimodal tasks.
Meanwhile, MMEF~\cite{denoeux2020representations} integrates Dempster-Shafer theory~\cite{denoeux2020representations} with a DL network to achieve reliable feature extraction. Nevertheless, existing approaches still struggle to fully leverage the complementary features of gastric MRI modalities, limiting their effectiveness in achieving reliable lesion segmentation.

\begin{figure*}[!t]
    \centering      
    \includegraphics[width=0.8\textwidth]{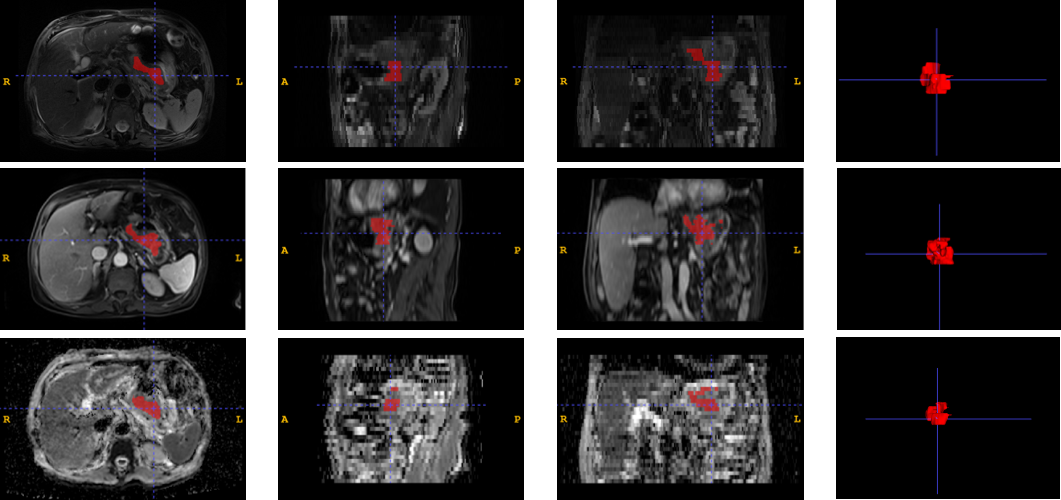}
    \caption{The data visualization for GCM 2025 dataset. From top to bottom, the images demonstrate the FS-T2W, CE-T1W, and ADC modalities.}
    \label{fig: GCM}
\end{figure*}

This study addresses challenges in multimodal MRI segmentation of gastric lesions with two principal contributions: (1) \textbf{GCM (Gastric Cancer MRI) 2025 dataset}, the first publicly available multimodal MRI resource for gastric cancer research, featuring fat-suppressed T2-weighted (FS-T2W), venous phase contrast-enhanced T1-weighted (CE-T1W), and apparent diffusion coefficient (ADC) images from 500 gastric cancer patients (as shown in Fig. \ref{fig: GCM}), is released to provide essential baseline data for lesion segmentation; and (2) \textbf{HWA-UNETR} (\textbf{H}ierarchical \textbf{W}indow \textbf{A}ggregate \textbf{UNE}t \textbf{TR}ansformer), a 3D multimodal segmentation framework is proposed, combining the novel HWA blocks for dynamic cross-modal feature correspondence and TFM blocks for global, multi-scale feature modeling through its innovative tri-orientated fusion mamba mechanism. Extensive experiments on multimodal benchmarks, including GCM 2025 and BraTS 2021 datasets, demonstrate the technical superiority of HWA-UNETR. The proposed framework achieves a Dice score improvement of up to 1.68\% compared to State-Of-The-Art (SOTA) methods, while maintaining exceptional robustness across various organ structures and multimodal imaging scenarios.

\section{GCM 2025: Gastric Cancer MRI Dataset}

\textbf{GCM} (\textbf{G}astric \textbf{C}ancer \textbf{M}RI Dataset) 2025 is currently recognized as the first large-scale open-source dataset featuring professionally annotated gastric cancer MRI examinations, integrating multimodal MRI results from 500 gastric cancer patients. It establishes standardized baselines for multimodal DL segmentation techniques, thereby advancing epidemiological research on gastric cancer.\\
\textbf{Dataset Modalities}: The standardization protocol of GCM 2025 includes FS-T2W, CE-T1W, and ADC images, as illustrated in Fig. \ref{fig: GCM}, where FS-T2W/CE-T1W demonstrate high soft-tissue contrast for lesion detection and tumor staging~\cite{bammer2003basic}, while ADC-based diffusion imaging noninvasively quantifies water molecular mobility to characterize tumor pathophysiology~\cite{liu2014apparent}.\\
\textbf{Dataset Construction}: MRI scans of GCM 2025 were acquired from November 2017 to December 2024. All examinations were performed using 1.5T or 3.0T scanners equipped with dedicated body coils. Sensitive patient information in this dataset has been removed. Each volume was annotated by a professional doctor and cross-checked by a second doctor. Detailed data acquisition information is provided at this public https 
\href{https://github.com/JeMing-creater/HWA-UNETR}{URL}.

\section{Methodology} 
\begin{figure*}[!t]
    \centering      
    \includegraphics[width=\textwidth]{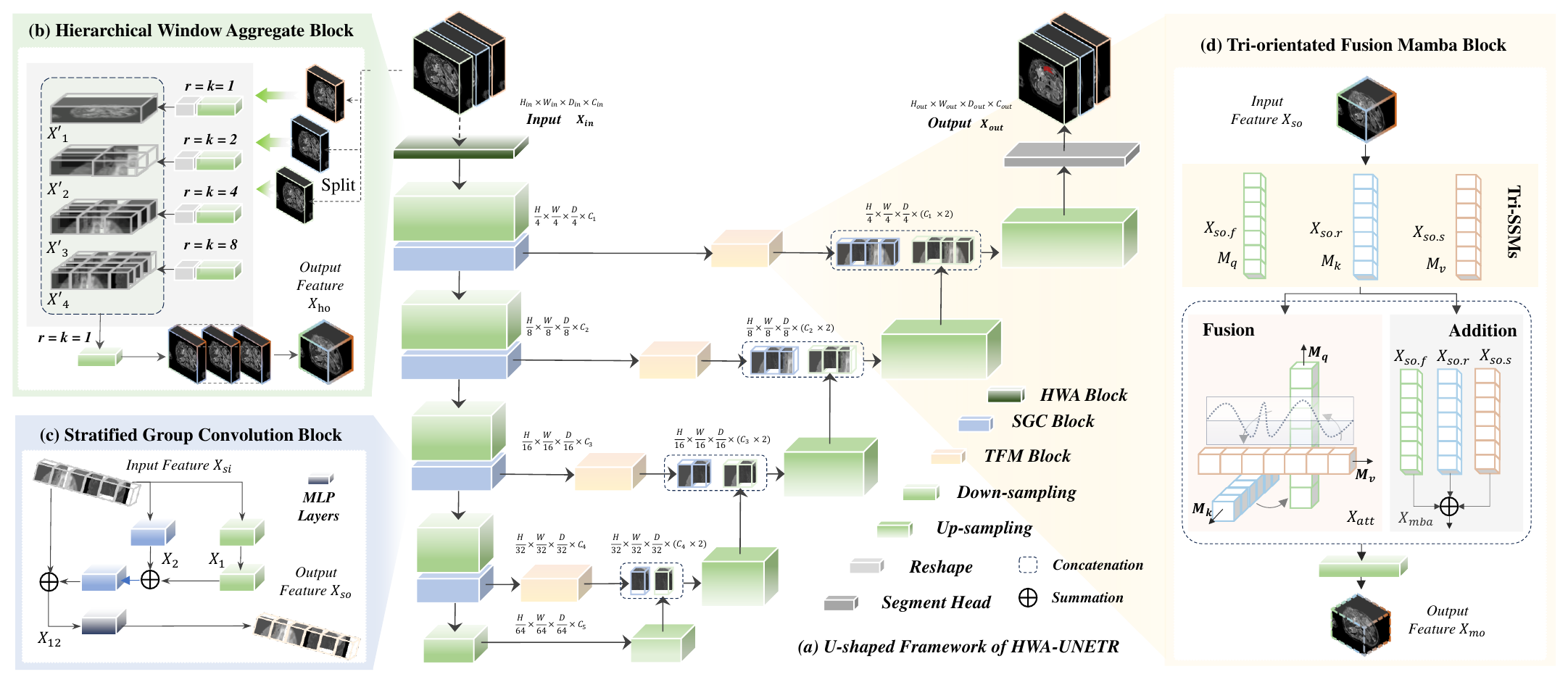}
    \caption{Overview of HWA-UNETR.}
    \label{fig: overview}
\end{figure*}

To address the challenges of multimodal gastric lesion segmentation, the HWA-UNETR framework is proposed, which achieves high-performance multimodal segmentation tasks through novel blocks and architecture design. The detailed implementation is illustrated in Fig. \ref{fig: overview} and described in the following sections.

\subsection{U-shaped Framework Overview}
HWA-UNETR adopts a 4-stage U-shaped architecture comprising an encoder-decoder structure with skip connections, which is shown in Fig. \ref{fig: overview} (a), aiming to achieve high-performance 3D multimodal image segmentation. The input volumetric data \( X_{in} \) with dimensions \(H_{in} \times W_{in} \times D_{in} \times C_{in}\) (height, width, depth, modals) undergoes hierarchical feature extraction in the encoder. At each stage \(i\), spatial resolution is progressively reduced to \(\frac{H}{2^{i+1}} \times \frac{W}{2^{i+1}} \times \frac{D}{2^{i+1}}\) via depthwise separable convolutional layers followed by instance normalization and ReLU activation. In the decoder, transposed convolutional layers upsample feature maps to recover the original spatial resolution. Simultaneously, skip connections integrate multi-scale encoder features to minimize information loss. Finally, a segmentation head equipped with a sigmoid activation layer produces pixel-wise probability maps, facilitating precise and scalable image segmentation.

To enhance its core functionality, HWA-UNETR integrates three key components within its framework:  \textbf{H}ierarchical \textbf{W}indow \textbf{A}ggregate (HWA) block,  \textbf{S}tratified \textbf{G}roup \textbf{C}onvolution (SGC) block, and  \textbf{T}ri-orientated \textbf{F}usion \textbf{M}amba (TFM) block. These components are introduced in the following sections.


\subsection{Hierarchical Window Aggregate Block}
The HWA block, whose structure is shown in Fig. \ref{fig: overview} (b), operates at the multimodal volumetric input $X_{in}$ stage, dynamic window partitioning, and multi-kernel convolutions are employed to align cross-modal local semantics. This structure allows for the extraction of features from multimodal images with cross-resolution characteristics, generating semantically consistent fused image features.
Specifically, the HWA Block adaptively employs depthwise convolutional layers with varying strides \( r \) and kernel sizes \( k \) based on the input image shapes to extract features from multimodal images, as detailed in Eq. \eqref{eq: hwa}.

\begin{equation}
\begin{aligned}
[X_{1}, X_{2},..., X_{C_{in}}] &= Sp(X_{in}); \\
X'_{i} = R(Dw(X_{i})_{1})\parallel R(Dw(X_{i})_{2})&\parallel R(Dw(X_{i})_{4}) \parallel R(Dw(X_{i})_{8});\\
X_{ho} = w_1 X'_{1}+ w_2 X'_{2} & + ...+  w_{C_{in}} X'_{C_{in}};\\
\end{aligned}
\label{eq: hwa}
\end{equation}
where $X_{ho}$ represents the output of the HWA Block; $+$ and $\parallel$ denote addition operation and concatenation operation performed along dimension $C_{in}$; $w_i$ represents the learnable parameters; $Dw(.)_{i}$ represents the depthwise convolutional layer with $r$ = $k$ = $i$; $R$ denotes the reshape operation that upsamples the convolution results back to their original size; and $Sp(.)$ represents a splitting operation that divides the input data into separate vectors via the $C_{in}$ dimension.

\subsection{Stratified Group Convolution Block}
The SGC block, deployed post-downsampling in each encoder stage, enhances local feature retention and spatial information preservation for cross-level fusion~\cite{xing2024segmamba}. As shown in Fig. \ref{fig: overview} (c), the tri-branch architecture processes shared input \(X_{si}\) via: (1) dual depthwise convolutions ($k$=3,1) with instance normalization obtain \(X_1\); (2) pointwise convolution refinement obtain \(X_2\); and (3) residual-enhanced fusion with MLP transformation obtain \(X_{so}\) as result of SGC block.

\subsection{Tri-orientated Fusion Mamba Block}
The TFM Block is deployed in the skip connections stage, specifically designed for modeling global and multi-scale features, as shown in Fig. \ref{fig: overview} (d).

In particular, to perform complementary feature modeling on cross-modal feature maps, the TFM block introduces a novel Tri-orientated Fusion Mamba computational mechanism, which integrates the long-sequence modeling capability of Tri-SSMs~\cite{xing2024segmamba} with the cross-modal attention mechanisms inherent in the attention~\cite {hatamizadeh2022swin} mechanism. This mechanism, as mathematically formulated in Eq. \eqref{eq: tfm}, provides enriched feature representations specifically optimized for the decoder's upsampling operations.

\begin{equation}
\begin{aligned}
M_q, M_k, M_v =  Ma(X_{so.f}&), Ma(X_{so.r}), Ma(X_{so.s});\\
X_{mba} = M_q& + M_k + M_v; \\
X_{att} = S (&\frac{M_qM_k^{T}}{\sqrt{d_{M_k}}} )M_v; \\
X_{mo}  = Dw&(X_{mba}||X_{att})_1;
\end{aligned}
\label{eq: tfm}
\end{equation}
where $X_{mo}$ denotes the output of TFM Block; $S$ denotes the softmax function~\cite{han2024agent}, \(Ma\) refers to the Mamba module that models global information within the sequence, \(X_{so.f}\), \(X_{so.r}\), and \(X_{so.s}\) denote the forward, reverse, and inter-slice direction feature maps, respectively.

\section{Experiments}
\subsection{Assistance Dataset}

\textbf{BraTS 2021} dataset~\cite{baid2021rsna} serves as an accessorial benchmark for evaluating HWA-UNETR's performance in multimodal MRI segmentation, particularly assessing its generalizability across tumor subregions. It contains 1,251 3D brain MRI cases, each comprising four imaging modalities (T1, T1Gd, T2, and T2-FLAIR) with annotations for three distinct tumor compartments: Whole Tumor (WT), Enhancing Tumor (ET), and Tumor Core (TC).

\subsection{Implementation Details}
Our model is implemented in PyTorch 2.4.1, CUDA 12.4, and MONAI 1.3.2, trained with random crops of 128×128×64 (GCM 2025) and 128×128×128 (BraTS 2021), batch size 2 per GPU. Both experiments utilized the AdamW optimizer for 300 epochs with an initial learning rate of 0.001, incorporating linear warm-up, cosine annealing, a weight decay of 0.4, and 1:1 positive-to-negative sample balancing, while employing a composite loss function that integrates soft dice loss ~\cite{milletari2016v} and focal loss ~\cite{lin2017focal}. All data underwent random axis flipping (50\% prob.) and intensity scaling/shifting (20\% prob.). Experiments run on a cloud platform with four NVIDIA A100 GPUs, randomly allocating 3D volumes into 70\% training, 10\% validation, and 20\% testing.

\subsection{Performance}

We compare HWA-UNETR with five SOTA segmentation methods: CNN-based nnUNet~\cite{isensee2020nnu}, Transformer-based Swin UNETR~\cite{hatamizadeh2022swin}, SSM-based SegMamba~\cite{xing2024segmamba} / nnMamba~\cite{gong2024nnmamba}, and MMEF~\cite{huang2025deep}, a method specifically designed for multimodal segmentation. All methods were retrained using their official implementations under consistent settings for fair comparison. Performance was quantitatively evaluated on GCM 2025 and BraTS 2021 datasets using Dice coefficient (Dice) and 95\% Hausdorff Distance (HD95).\\
\textbf{GCM 2025} : 

\begin{table*}[!t]
\centering
\caption{
Comparative Performance Analysis of Various Network Architectures for Lesion Segmentation on the GCM 2025 Dataset. The performance metrics of HWA-UNETR are highlighted in \colorbox[HTML]{EFEFEF}{gray}, whereas the best segmentation performance is marked in \textcolor[HTML]{9A0000}{\textbf{red}}.
}
\resizebox{\linewidth}{!}{
\begin{tabular}{c|cccc|c}
\hline
                         & \multicolumn{4}{c|}{Dice Similarity Coefficient (\%)}                                                                                                                                                           &                                                  \\ \cline{2-5}
\multirow{-2}{*}{Method}                                                            & FS-T2W↑                                    & CE-T1W↑ & ADC↑                                &\textbf{ Avg.}↑                                                         & \multirow{-2}{*}{\makecell{Average  Hausdorff \\ Distance 95\% ($mm$) ↓}} \\ \hline
nnUNet~\cite{isensee2020nnu}                                                                            & 70.47                                   & 71.68       & 67.31                        & 69.82                                                         & 6.54                                             \\
Swin-UNETR~\cite{hatamizadeh2022swin}                                                                        & 72.06                                   & 72.46  & 69.83                             & 71.45                                                         & 5.67                                             \\
SegMamba~\cite{xing2024segmamba}                                                                          & 71.6                                   & 73.83  & 71.05                             & 72.16                                                         & 5.37                                             \\
nnMamba~\cite{gong2024nnmamba}                                                                        & 70.12                                   & 73.02   & 69.94                            & 71.03                                                        & 5.86                                             \\
MMEF-nnUNet~\cite{huang2025deep}                                                                       & 72.66   & 73.69  & 71.24                             & 72.53                                                         & \color[HTML]{680100} \textbf{5.03}                                          \\ \hline
\rowcolor[HTML]{EFEFEF} 
HWA-UNETR(ours)       & \color[HTML]{680100} \textbf{74.76}                                  & \cellcolor[HTML]{EFEFEF}{\color[HTML]{680100} \textbf{75.56}} & {\color[HTML]{680100} \textbf{72.31}} & \cellcolor[HTML]{EFEFEF}{\color[HTML]{680100} \textbf{74.21}} & 5.35            \\ \hline
\end{tabular}
}
\label{tab:gcm_a}
\end{table*}

\begin{figure*}[!t]
    \centering      
    \includegraphics[width=\textwidth]{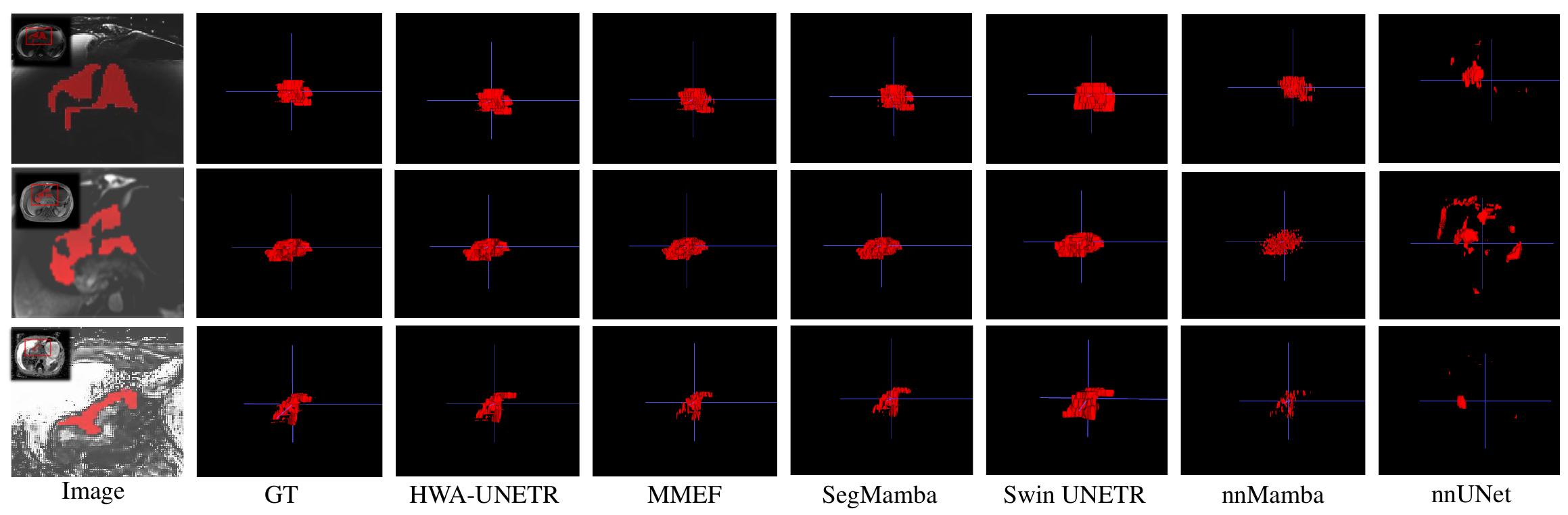}
    \caption{Visual comparisons of proposed HWA-UNETR and other SOTA methods. The arrangement of the images in this figure corresponds to that in Fig. \ref{fig: GCM}.}
    \label{fig: Visual}
\end{figure*}

\begin{table*}[!t]
\centering
\caption{
Comparative Performance Analysis of Various Network Architectures for Lesion Segmentation on the BraTS 2021 Dataset. The colors of the record in this table are labeled the same as in Table \ref{tab:gcm_a}.
}
\begin{tabular}{c|cccc|c}
\hline
                         & \multicolumn{4}{c|}{Dice Similarity Coefficient (\%)}                                                                                                                                                           &                                                  \\ \cline{2-5}
\multirow{-2}{*}{Method}                                                            & WT↑                                   & ET↑  & TC↑                                 &\textbf{ Avg.}↑                                                         & \multirow{-2}{*}{\makecell{Average  Hausdorff \\ Distance 95\% ($mm$) ↓}} \\ \hline
nnUNet~\cite{isensee2020nnu}                                                                            & 92.72                                 & 88.34  & 91.39                               & 90.84                                                         & 5.33                                             \\
Swin-UNETR~\cite{hatamizadeh2022swin}                                                                        & 93.32                                 & 89.08  & 91.69                               & 91.36                                                         & 5.03                                             \\
SegMamba~\cite{xing2024segmamba}                                                                          & 93.13                                 & 88.21  & 93.82                               & 91.72                                                         & 4.73                                             \\
nnMamba~\cite{gong2024nnmamba}                                                                        & 92.52                                 & 86.68  & 93.89                               & 91.03                                                        & 5.33                                             \\
MMEF-nnUNet~\cite{huang2025deep}                                                                       & {\color[HTML]{680100} \textbf{93.58}} & 89.04  & 93.77                               & 92.13                                                         & 3.97                                             \\ \hline
\rowcolor[HTML]{EFEFEF} 
HWA-UNETR(ours)       & 93.41                                 & {\color[HTML]{680100} \textbf{89.54}} & \cellcolor[HTML]{EFEFEF}{\color[HTML]{680100} \textbf{94.07}} & \cellcolor[HTML]{EFEFEF}{\color[HTML]{680100} \textbf{92.34}} & {\color[HTML]{680100} \textbf{3.51}}             \\ \hline
\end{tabular}
\label{tab:bra_a}
\end{table*}
Table \ref{tab:gcm_a} summarizes the multimodal segmentation results of HWA-UNETR and other SOTA methods on the newly released GCM 2025 dataset. The outstanding segmentation performance demonstrated in the results validates the effectiveness of our approach. Although HWA-UNETR ranks second only to the method MMEF~\cite{huang2025deep} in terms of the optimal segmentation edge metric HD95 (5.03 mm), it achieves the highest average Dice score of 74.21\%, surpassing other SOTA methods by 1.68\%. This performance highlights HWA-UNETR's superior capability in handling multimodal segmentation tasks with misaligned information.\\
\textbf{BraTS 2021} :
The results on the BraTS 2021 dataset are listed in Table \ref{tab:bra_a}. The proposed HWA-UNETR achieves optimal performance compared to other methods, achieving an average Dice score of 92.34\% along with an average HD95 of 3.51 mm. These compelling results further substantiate two key advantages of HWA-UNETR: (1) its advanced feature integration capabilities across diverse imaging modalities, and (2) its remarkable robustness in cross-organ segmentation tasks. \\
\textbf{Visual Comparisons} :
Fig. \ref{fig: Visual} provides an intuitive comparison of the segmentation results obtained by different methods. On the GCM 2025 dataset, our HWA-UNETR accurately detects the boundaries of tumor regions in each modality image, demonstrating better consistency compared to other SOTA methods.

\subsection{Ablation Study}
Table \ref{tab:ab} validates the effectiveness of core modules in the HWA-UNETR framework. On the GCM 2025 dataset: (1) The TFM Block establishes excellent performance (i.e, 71.84\% Dice score) through global context modeling and hierarchical feature integration; (2) the SGC Block enhances Dice score to 72.47\% by capturing local information; and (3) the HWA Block achieves a 1.85\% Dice score improvement (74.21\%) via dynamic window partitioning strategy, demonstrating superior capability in handling misaligned multimodal data.

\begin{table*}[!t]
\caption{Ablation study of various block settings of HWA-UNETR on the GCM 2025 benchmark. The colors of the record in this table are labeled the same as in Table \ref{tab:gcm_a}.}
\centering
\scalebox{1.0}{
\begin{tabular}{ccc|c|c}
\hline
HWA & SGC & TFM & Avg. Dice↑                             & Avg. HD 95 ↓                           \\ \hline
×   & ×   & ×   & 67.56                                 & 10.57                                 \\
×   & ×   & \checkmark   & 71.84                                 & 6.03                                  \\
×   & \checkmark   & \checkmark   & 72.36                                 & \color[HTML]{680100} \textbf{5.31}                                  \\ \hline
\rowcolor[HTML]{EFEFEF} 
\checkmark   & \checkmark   & \checkmark   & {\color[HTML]{680100} \textbf{74.21}} & {5.35} \\ \hline
\end{tabular}
}
\label{tab:ab}
\end{table*}

\section{Conclusion}
In this paper, HWA-UNETR is proposed as a novel multimodal 3D medical image segmentation framework specifically tailored for gastric cancer lesions. The framework designs a hierarchical window aggregation HWA Block with a learnable cross-modal alignment layer to address the structural differences between unaligned MRI modalities. In addition, a new tri-orientated fusion mamba is introduced to model global and multi-scale features for multimodal data specifically.  To advance gastric cancer research, GCM 2025, the first large-scale open-source multimodal MRI dataset, is publicly released, comprising 500 expert-annotated, aligned multimodal MRI scans (FS-T2W/CE-T1W/ADC) for gastric cancer segmentation. Extensive experiments on both the newly released GCM 2025 and the public BraTS 2021 datasets demonstrate that HWA-UNETR achieves SOTA segmentation accuracy on gastric multimodal MRI scans while exhibiting exceptional generalizability in cross-organ segmentation tasks, outperforming conventional approaches by a significant margin.

\begin{credits}
\subsubsection{\ackname} 
This work was supported in part by the National Key Research and Development Program of China (2022ZD0117700), the National Natural Science Foundation of China (U21A20520, 62325204), the Key-Area Research and Development Program of Guangzhou City (2023B01J1001), and the Hainan University High-level Talent Research Launch Fund (XJ2400012551).

\subsubsection{\discintname}
This study and its authors have no competing interests.
\end{credits}

\bibliographystyle{splncs04}
\bibliography{Paper-3634}

\begin{thebibliography}{10}
\providecommand{\url}[1]{\texttt{#1}}
\providecommand{\urlprefix}{URL }
\providecommand{\doi}[1]{https://doi.org/#1}

\bibitem{ali2024review}
Ali, M., Wu, T., Hu, H., Luo, Q., Xu, D., Zheng, W., Jin, N., Yang, C., Yao, J.: A review of the segment anything model (sam) for medical image analysis: Accomplishments and perspectives. Computerized Medical Imaging and Graphics p. 102473 (2024)

\bibitem{baid2021rsna}
Baid, U., Ghodasara, S., Mohan, S., Bilello, M., Calabrese, E., Colak, E., Farahani, K., Kalpathy-Cramer, J., Kitamura, F.C., Pati, S., et~al.: The rsna-asnr-miccai brats 2021 benchmark on brain tumor segmentation and radiogenomic classification. arXiv preprint arXiv:2107.02314  (2021)

\bibitem{bammer2003basic}
Bammer, R.: Basic principles of diffusion-weighted imaging. European Journal of Radiology  \textbf{45}(3),  169--184 (2003)

\bibitem{bibars2023cross}
Bibars, M., Salah, P.E., Eldeib, A., Elattar, M.A., Yassine, I.A.: Cross-modality deep transfer learning: Application to liver segmentation in ct and mri. In: Annual Conference on Medical Image Understanding and Analysis. pp. 96--110. Springer (2023)

\bibitem{borggreve2019imaging}
Borggreve, A.S., Goense, L., Brenkman, H.J., Mook, S., Meijer, G.J., Wessels, F.J., Verheij, M., Jansen, E.P., Van~Hillegersberg, R., Van~Rossum, P.S., et~al.: Imaging strategies in the management of gastric cancer: current role and future potential of mri. The British Journal of Radiology  \textbf{92}(1097),  20181044 (2019)

\bibitem{denoeux2020representations}
Den{\oe}ux, T., Dubois, D., Prade, H.: Representations of uncertainty in artificial intelligence: Probability and possibility. A Guided Tour of Artificial Intelligence Research: Volume I: Knowledge Representation, Reasoning and Learning pp. 69--117 (2020)

\bibitem{gong2024nnmamba}
Gong, H., Kang, L., Wang, Y., Wan, X., Li, H.: nnmamba: 3d biomedical image segmentation, classification and landmark detection with state space model. arXiv preprint arXiv:2402.03526  (2024)

\bibitem{han2024agent}
Han, D., Ye, T., Han, Y., Xia, Z., Pan, S., Wan, P., Song, S., Huang, G.: Agent attention: On the integration of softmax and linear attention. In: European Conference on Computer Vision. pp. 124--140. Springer (2024)

\bibitem{hatamizadeh2022swin}
Hatamizadeh, A., Nath, V., Tang, Y., Yang, D., Roth, H.R., Xu, D.: Swin unetr: Swin transformers for semantic segmentation of brain tumors in mri images. In: International MICCAI Brainlesion Workshop. pp. 272--284. Springer (2022)

\bibitem{huang2025deep}
Huang, L., Ruan, S., Decazes, P., Den{\oe}ux, T.: Deep evidential fusion with uncertainty quantification and reliability learning for multimodal medical image segmentation. Information Fusion  \textbf{113},  102648 (2025)

\bibitem{isensee2020nnu}
Isensee, F., J{\"a}ger, P.F., Full, P.M., Vollmuth, P., Maier-Hein, K.H.: nnu-net for brain tumor segmentation. In: International MICCAI Brainlesion Workshop. pp. 118--132. Springer (2020)

\bibitem{liang2023agilenet}
Liang, J., Huang, T., Li, D., Ding, Z., Li, Y., Huang, L., Wang, Q., Zhang, X.: Agilenet: A rapid and efficient breast lesion segmentation method for medical image analysis. In: Chinese Conference on Pattern Recognition and Computer Vision (PRCV). pp. 419--430. Springer (2023)

\bibitem{liang2024comprehensive}
Liang, J., Zhang, M., Tan, C., Huang, T., Zhang, X., Zhang, Z., Gao, S., Sheng, Q., Pang, Y.: Comprehensive transformer integration network (ctin): Advancing endoscopic disease segmentation with hybrid transformer architecture. In: Chinese Conference on Pattern Recognition and Computer Vision (PRCV). pp. 210--224. Springer (2024)

\bibitem{lin2017focal}
Lin, T.Y., Goyal, P., Girshick, R., He, K., Doll{\'a}r, P.: Focal loss for dense object detection. In: Proceedings of the IEEE International Conference on Computer Vision. pp. 2980--2988 (2017)

\bibitem{liu2014apparent}
Liu, S., Guan, W., Wang, H., Pan, L., Zhou, Z., Yu, H., Liu, T., Yang, X., He, J., Zhou, Z.: Apparent diffusion coefficient value of gastric cancer by diffusion-weighted imaging: correlations with the histological differentiation and lauren classification. European Journal of Radiology  \textbf{83}(12),  2122--2128 (2014)

\bibitem{ma2024segment}
Ma, J., He, Y., Li, F., Han, L., You, C., Wang, B.: Segment anything in medical images. Nature Communications  \textbf{15}(1), ~654 (2024)

\bibitem{milletari2016v}
Milletari, F., Navab, N., Ahmadi, S.A.: V-net: Fully convolutional neural networks for volumetric medical image segmentation. In: 2016 Fourth International Conference on 3D Vision (3DV). pp. 565--571. IEEE (2016)

\bibitem{pang2025online}
Pang, Y., Li, Y., Huang, T., Liang, J., Wang, Z., Dong, C., Kuang, D., Hu, Y., Chen, H., Lei, T., et~al.: Online self-distillation and self-modeling for 3d brain tumor segmentation. IEEE Journal of Biomedical and Health Informatics  (2025)

\bibitem{pang2023slim}
Pang, Y., Liang, J., Huang, T., Chen, H., Li, Y., Li, D., Huang, L., Wang, Q.: Slim unetr: Scale hybrid transformers to efficient 3d medical image segmentation under limited computational resources. IEEE Transactions on Medical Imaging  (2023)

\bibitem{ronneberger2015u}
Ronneberger, O., Fischer, P., Brox, T.: U-net: Convolutional networks for biomedical image segmentation. In: International Conference on Medical image computing and computer-assisted intervention. pp. 234--241. Springer (2015)

\bibitem{shi2023h}
Shi, J., Kan, H., Ruan, S., Zhu, Z., Zhao, M., Qiao, L., Wang, Z., An, H., Xue, X.: H-denseformer: An efficient hybrid densely connected transformer for multimodal tumor segmentation. In: International Conference on Medical Image Computing and Computer-Assisted Intervention. pp. 692--702. Springer (2023)

\bibitem{singh2024treatment}
Singh, A., Adams-Tew, S., Johnson, S., Odeen, H., Shea, J., Johnson, A., Day, L., Pessin, A., Payne, A., Joshi, S.: Treatment efficacy prediction of focused ultrasound therapies using multi-parametric magnetic resonance imaging. In: MICCAI Workshop on Cancer Prevention through Early Detection. pp. 190--199. Springer (2024)

\bibitem{sung2021global}
Sung, H., Ferlay, J., Siegel, R.L., Laversanne, M., Soerjomataram, I., Jemal, A., Bray, F.: Global cancer statistics 2020: Globocan estimates of incidence and mortality worldwide for 36 cancers in 185 countries. CA: A Cancer Journal for Clinicians  \textbf{71}(3),  209--249 (2021)

\bibitem{wenxuan2021transbts}
Wenxuan, W., Chen, C., Meng, D., Hong, Y., Sen, Z., Jiangyun, L.: Transbts: Multimodal brain tumor segmentation using transformer. In: International Conference on Medical Image Computing and Computer-Assisted Intervention, Springer. pp. 109--119 (2021)

\bibitem{xing2024segmamba}
Xing, Z., Ye, T., Yang, Y., Liu, G., Zhu, L.: Segmamba: Long-range sequential modeling mamba for 3d medical image segmentation. In: International Conference on Medical Image Computing and Computer-Assisted Intervention. pp. 578--588. Springer (2024)

\bibitem{zhou2019review}
Zhou, T., Ruan, S., Canu, S.: A review: Deep learning for medical image segmentation using multi-modality fusion. Array  \textbf{3},  100004 (2019)

\end{thebibliography}

\end{document}